\documentclass[12pt]{article}
\usepackage{amsfonts}
\usepackage{myart}

\input tcilatex
\begin{document}

\title{Necessity of the general relativity revision and free motion of
particles in non-Riemannian space-time geometry}
\author{Yuri A.Rylov}
\date{Institute for Problems in Mechanics, Russian Academy of Sciences,\\
101-1, Vernadskii Ave., Moscow, 119526, Russia.\\
e-mail: rylov@ipmnet.ru\\
Web site: {$http://rsfq1.physics.sunysb.edu/\symbol{126}rylov/yrylov.htm$}\\
or mirror Web site: {$http://gasdyn-ipm.ipmnet.ru/\symbol{126}%
rylov/yrylov.htm$}}
\maketitle

\begin{abstract}
It is shown, that a free motion of microparticles (elementary particles) in
the gravitational field is multivariant (stochastic). This multivariance is
conditioned by multivariant physical space-time geometry. The physical
geometry is described completely by a world function. The Riemannian
geometries form a small part of possible physical geometries. The
contemporary theory of gravitation ignores existence of physical geometries.
It supposes, that any space-time geometry is a Riemannian geometry. It is a
mistake. As a result the contemporary theory of gravitation needs a
revision. Besides, the Riemannian geometry is inconsistent, and conclusions
of the gravitational theory, based on inconsistent geometry may be invalid.
Free motion of macroparticles (planets), consisting of many connected
microparticles, is deterministic, because connection of microparticles
inside the macroparticle averages stochastic motion of single microparticles.
\end{abstract}


\section{Introduction}

The physical geometry describes the space-time in terms of a finite
space-time distance $\rho $ (or in terms of world function $\sigma =\frac{1}{%
2}\rho ^{2}$). This description in terms of distance $\rho $ and only in
terms of $\rho $ is a complete description. The Riemannian geometry tries to
describe the space-time in terms of infinitesimal space-time distance $d\rho 
$, which is determined by the metric tensor $g_{ik}$, given at any point of
the space-time. 
\begin{equation}
d\rho ^{2}=g_{ik}\left( x\right) dx^{i}dx^{k}  \label{a0.1}
\end{equation}%
The distance $\rho =\rho \left( x,x^{\prime }\right) $ is a function of two
space-time points $x$ and $x^{\prime }$. The distance $\rho $ as a function
of two points contains much more information, than ten functions $g_{ik}$,
which are functions of one point. To obtain the finite distance $\rho $ from
the infinitesimal distance $d\rho $, a set of additional conditions (and
additional information) is to be fulfilled. In particular, the obtained 
\textit{finite distance }$\rho $\textit{\ is to be a single-valued function
of any two points of the space-time}. Construction of the space-time
geometry by methods of the Riemannian geometry construction leads, in
general, to a many-valued finite distance $\rho $. This fact is a nonsense.
Besides, the Riemannian geometry in itself appeared to be inconsistent \cite%
{R2005,R2006}. However, in the twentieth century nobody paid attention to
this fact, because there were no alternative to the Riemannian geometry.

When this alternative appeared \cite{R2007}, the problem of the general
relativity revision arose. Let us stress, that it was just the problem of
revision, but not a problem of a new gravitational theory construction,
because the main idea of the general relativity (geometrization of physics)
remains changeless.

The problem of the general relativity revision contains two essential points

1. A use of more effective and general mathematical method: the physical
geometry, described completely by the finite space-time distance and only by
it.

2. Description of the relativity theory in terms, which are adequate to this
theory.

In fact, when the relativity theory came in the stead of the nonrelativistic
physics, some terms and concepts of the nonrelativistic physics remained in
the relativity theory. These concepts do not prevent one to solve concrete
physical problems. However, they prevent from development of the relativity
theory.\ In particular, nonrelativistic concept of two events nearness
prevents from the general relativity generalization \cite{R2009}.

The present paper is written in the framework of the physics geometrization.
The program of the physics geometrization is a generalization of the
relativity theory on the case of non--Riemannian space-time geometry. The
general relativity (theory of gravitation) is created at the supposition,
that the space-time geometry is a Riemannian geometry. In the twentieth
century the Riemannian geometry was considered as the most general kind of
geometry, which is available for the space-time description. However, it
appears, that there exist non-Riemannian geometries, which are more general
in the sense, that the set of Riemannian geometries is only a negligible
part of the set of non-Riemannian (physical) geometries. Physical geometries
are described completely by a finite space-time interval, but not by
infinitesimal space-time interval as Riemannian geometries. Physical
geometry is a constructive (nonaxiomatizable) geometry, which cannot be
deduced from axiomatics.

It has been shown \cite{R2009}, that a heavy sphere creates a deformation of
the space-time in such a way, that the space-time geometry ceases to be a
Riemannian geometry. The fact is that, the metric tensor, given in the whole
space-time, determines the space-time geometry only under condition that the
space-time geometry is Riemannian. In the twentieth century the Riemannian
geometry was considered as a most general space-time geometry. The
condition, that the space-time geometry is a Riemannian geometry, seemed to
be very reasonable.

The Riemannian geometry is a mathematical geometry in the sense, that it is
a logical construction, and all propositions of mathematical geometry are
deduced from a system of axioms. The main property of a mathematical
geometry is the fact, that any mathematical geometry is a logical
construction. It is a secondary circumstance, whether or not the
mathematical geometry describes mutual disposition of geometrical objects.
There are such mathematical geometries (for instance, symplectic geometry),
which do not describe a disposition of geometrical objects.

On the contrary, any physical geometry is defined as a science on mutual
disposition of geometrical objects in the space or in the space-time. It is
a secondary circumstance, whether or not the physical geometry is a logical
construction. The physical geometry is described completely by means of a
distance $\rho \left( P,Q\right) $ between any two points $P$ and $Q$ of the
space, or of the space-time. In this sense the physical geometry is a
distance geometry \cite{B1953}. Physical geometry distinguishes from the
distance (metric) geometry in the sense, that the distance geometry is not
completely a metric geometry. For instance, in the distance geometry the
concept of a curve is formulated not only in terms of a distance, whereas in
physical geometry all geometrical concepts are formulated in terms of a
distance $\rho $ and only in terms of distance. In fact the physical
geometry is formulated in terms of the world function $\sigma \left(
P,Q\right) =\frac{1}{2}\rho ^{2}\left( P,Q\right) $ \cite{S1960}. A use of
world function is more effective from technical viewpoint. (For instance,
the scalar product $\left( \mathbf{PQ}.\mathbf{RS}\right) $ of vectors $%
\mathbf{PQ}$ and $\mathbf{RS}$ is a linear function of world functions of
points $P,Q,R,S$, whereas in terms of the distance $\rho $ the scalar
product is a more complicated expression.

In reality there exist nonaxiomaitzable space-time geometries, which cannot
be deduced from a system of axioms. Riemannian geometries form a small
subset of all possible physical space-time geometries, which are
nonaxiomatizable, in general. Different nonaxiomatizable space-time
geometries may have the same metric tensor. As a result metric tensor does
not determine the space-time geometry uniquely.

Besides, the Riemannian geometry is inconsistent, in general. The Riemannian
geometry is considered as a kind of a mathematical geometry, i.e. it can be
deduced from a system of axioms. These axioms are inconsistent at some
points. As a result the Riemannian have at least three well known defects.
First, a parallel transport of a vector from the point $P_{0}$ to the point $%
P_{1}$ depends on the path of the transport. It means that there is no
absolute parallelism in the Riemannian geometry. Second, from physical
viewpoint the main characteristic of the space-time geometry is a distance $%
\rho \left( P,Q\right) $ between two space-time points $P$ and $Q$. This
distance must be single-valued. However, in the Riemannian geometry the
distance $\rho \left( P,Q\right) $ is defined as the length of the geodesic,
connecting points $P$ and $Q$. In the Riemannian geometry there are such
points $P$ and $Q$ , which my be connected by several geodesics of different
length. It leads to multiformity of the distance $\rho \left( P,Q\right) $.

Third, the Euclidean geometry is a partial case of a Riemannian geometry.
Let us construct Euclidean geometry on a two-dimensional plane $\mathcal{P}%
_{2}$ by means of the method of the Riemannian geometry construction, i.e.
we define the distance $\rho \left( P,Q\right) $ as a length of a geodesic
(the shortest line, connecting points $P$ and $Q$). As far as in the case of
Euclidean geometry the geodesics are straight lines, one obtains in the
Cartesian coordinate system

\begin{equation}
\rho \left( P,Q\right) =\rho \left( \mathbf{x},\mathbf{x}^{\prime }\right) =%
\sqrt{\left( \mathbf{x-x}^{\prime }\right) ^{2}}  \label{a1.1}
\end{equation}%
Let us cut a hole in the two-dimensional plane $\mathcal{P}_{2}$.The
geodesics, passing through the hole, become to be impossible, and the
shortest lines pass around the hole. Their lengths change, and the formula (%
\ref{a1.1}) ceases to be valid for some points. As a result the plane $%
\mathcal{P}_{2}$ with a hole cannot be embedded isometrically into the same
plane $\mathcal{P}_{2}$ without the hole.

This result is a paradox, because, if experimentally one cuts a hole in a
flat piece of tin-plate, the obtained piece with a hole can be embedded
isometrically in the original piece of tin-plate. Mathematicians know this
paradoxical result, which means, that the conventional method of the
Riemannian geometry construction is inconsistent. However, they have no
alternative to the conventional method of the Riemannian geometry
construction. As a result they prefer to consider Riemannian geometries on
convex sets of points.

Of course, such a restriction by the convex point sets does not solve the
problem of the Riemannian geometry inconsistency. This problem may be solved
only by a change of the geometry construction method.

Deduction of the geometrical propositions from a system of basic axioms is a
very laborious process. One needs to prove numerous theorems. Besides, one
should be sure that the basic axioms are compatible between themselves. A
test of this compatibility is a very laborious process. For any new
space-time geometry one needs to repeat this test of the geometry
consistency.

However, the main problem of the mathematical geometries construction is a
doubt, that any space-time geometry may be deduced from a finite system of
basic axioms. Indeed, any geometry is a continual set of geometrical
propositions. It follows from no quarter, that a continual set of
geometrical propositions can be deduced from a finite number of basic
propositions by means of the formal logic. It is true, that Euclid succeeded
to deduce all propositions of the Euclidean geometry from several axioms.
However, it does not mean that such a deduction is possible for other
geometries. In reality such a deduction is impossible for most of physical
geometries, i.e. for geometries, which can be used for the space-time
description \cite{R2005,R2006}.

For construction of physical geometries one should use the proper Euclidean
geometry itself (but not the method of its construction). The proper
Euclidean geometry $\mathcal{G}_{\mathrm{E}}$ is a mathematical geometry and
a physical geometry at the same time. It means, that the proper Euclidean
geometry $\mathcal{G}_{\mathrm{E}}$ can be deduced from a system of basic
axioms, and all its propositions can be expressed in terms of the world
function $\sigma _{\mathrm{E}}$ of the proper Euclidean geometry $\mathcal{G}%
_{\mathrm{E}}$. Replacing $\sigma _{\mathrm{E}}$ in all propositions of the
Euclidean geometry $\mathcal{G}_{\mathrm{E}}$ by the world function $\sigma $
of other physical geometry $\mathcal{G}$, one obtains all propositions of
the physical geometry $\mathcal{G}$.

The procedure of the world function replacement is a change of distances
between the points of the space (or space-time). Such a change is a
deformation of the Euclidean geometry. This method of a physical geometry
construction is called the deformation principle \cite{R2007}. The
deformation principle is very simple. It does not need a proof of numerous
theorems and a test of the geometry consistency. (All theorems has been
proved at the construction of the proper Euclidean geometry). A use of the
deformation principle for the physical geometry construction does not need
an application of the formal logic. The deformation principle admits one to
construct nonaxiomatizable geometries. Most of physical geometries are
nonaxiomaitzable, and they cannot be constructed by the conventional method
(deduction of a geometry from axioms). In particular, the Riemannian
geometry ($\sigma $-Riemannian one), constructed by means of the deformation
principle, has not defects, which are characteristic for the Riemannian
geometry, constructed by the conventional method. There is a
fern-parallelism in the $\sigma $-Riemannian geometry. The world function is
single-valued. Cutting a hole in the space-time, one does not change the
space-time geometry in the remaining part of the space-time.

Contemporary theory of gravitation as well as the contemporary cosmology are
based on the supposition, that the space-time geometry is a Riemannian
geometry. However, generalization of the general relativity on the case of a
physical space-time geometry shows that the deformation of the space-time
geometry of Minkowski generates a non-Riemannian space-time geometry \cite%
{R2009}. The obtained generalization admits one to obtain the world function
of the space-time geometry directly \cite{R2009}. In particular, the
space-time geometry, generated by a heavy sphere is non-Riemannian.

The world function of a Riemannian geometry satisfies the equation%
\begin{equation}
\sigma _{,i}\left( x,x^{\prime }\right) g^{ik}\left( x\right) \sigma
_{,k}\left( x,x^{\prime }\right) =2\sigma \left( x,x^{\prime }\right)
,\qquad \sigma _{,i}\left( x,x^{\prime }\right) \equiv \frac{\partial }{%
\partial x^{i}}\sigma \left( x,x^{\prime }\right)  \label{a1.2}
\end{equation}%
The first approximation of the world function of the space-time, generated
by the heavy sphere of mass $M$ has the form%
\begin{equation}
\sigma \left( t_{1},\mathbf{x}_{1};t_{2},\mathbf{x}_{2}\right) =\frac{1}{2}%
\left( c^{2}\left( 1-\frac{4GM}{c^{2}\left\vert \mathbf{x}_{2}\mathbf{+x}%
_{1}\right\vert }\right) \left( t_{2}-t_{1}\right) ^{2}-\left( \mathbf{x}_{2}%
\mathbf{-x}_{1}\right) ^{2}\right)  \label{a1.3}
\end{equation}%
where $G$ is the gravitational constant. The world function (\ref{a1.3})
does not satisfy equation (\ref{a1.2}). It describes non-Riemannian
space-time geometry, although the metric tensor coincides with the metric
tensor of Newtonian approximation%
\begin{equation}
g_{00}\left( \mathbf{x}\right) =c^{2}-2\frac{Gm}{\left\vert \mathbf{x}%
\right\vert },\qquad g_{0\alpha }=g_{\alpha \beta }=0,\qquad \alpha ,\beta
=1,2,3  \label{a1.4}
\end{equation}%
Note, that in Riemannian geometry, constructed for the metric tensor (\ref%
{a1.4}), the world function is many-valued, whereas the function of
non-Riemannian geometry is single-valued \cite{R2009}, as it follows from (%
\ref{a1.3}). This fact tells in behalf of non-Riemannian geometry.

Note, that a use of physical (non-Riemannian) geometry is not a hypothesis,
which should be tested by experiments. It is a logical necessity, because
the Riemannian geometry is incosistent. It means that contemporary theory of
gravitation and cosmology need a revision. We do not state, that such a
revision will lead to a change of our cosmological conceptions. However,
such a change may take place. For instance, the concept of dark matter, made
on the basis of unsatisfactory theory of gravitation may appear to be
invalid.

In this paper we try to obtain the law of free particle motion in
non-Riemannian space-time geometry. This law has been formulated for
microparticles, moving in the arbitrary space-time geometry \cite{R2009a}.
This law has been presented in invariant form (in terms of the world
function). Now we present it in terms of differential equations (in the
conventional coordinate form).

\section{Motion of a free microparticle in physical space-time geometry}

The state of a simple microparticle is described by its skeleton $\mathcal{P}%
^{1}$, which consists of two points $P_{0},P_{1}$. These two points form a
vector $\mathbf{P}_{0}\mathbf{P}_{1}=\left\{ P_{0},P_{1}\right\} $, which
describes the energy-momentum of the microparticle.\ The length $\left\vert 
\mathbf{P}_{0}\mathbf{P}_{1}\right\vert $ of vector $\mathbf{P}_{0}\mathbf{P}%
_{1}$ is the geometrical mass $\mu $ of the microparticle 
\begin{equation}
\mu =\left\vert \mathbf{P}_{0}\mathbf{P}_{1}\right\vert =\sqrt{2\sigma
\left( P_{0},P_{1}\right) }  \label{a2.1}
\end{equation}%
where $\sigma $ is the world function of space-time geometry. The
geometrical mass $\mu $ is connected with the usual mass $m$ of the
microparticle by means of relation%
\begin{equation}
m=b\mu  \label{a2.2}
\end{equation}%
where $b$ is some universal constant. There are complex microparticles,
whose skeleton $\mathcal{P}^{n}$ consists of $n+1$, $n=1,2,..$ space-time
points.

The world function of the space-time of Minkowski has the form%
\begin{equation}
\sigma _{\mathrm{M}}\left( P,P^{\prime }\right) =\sigma _{\mathrm{M}}\left(
x,x^{\prime }\right) =\frac{1}{2}g_{ik}\left( x^{i}-x^{\prime i}\right)
\left( x^{k}-x^{\prime k}\right)  \label{a2.3}
\end{equation}%
\begin{equation}
g_{ik}=\text{diag}\left\{ c^{2},-1,-1,-1\right\}  \label{a2.4}
\end{equation}%
Evolution of the microparticle in the space-time is described by a world
chain $\mathcal{T}_{\mathrm{br}}$ of connected links 
\begin{equation}
\mathcal{T}_{\mathrm{br}}=\dbigcup\limits_{s}\mathcal{P}_{s}^{1}=\dbigcup%
\limits_{s}\mathcal{T}_{\left[ s,s+1\right] }  \label{a2.5}
\end{equation}%
where any link $\mathcal{T}_{\left[ s,s+1\right] }$ is a set of points $R$,
defined by the relation 
\begin{equation}
\mathcal{T}_{\left[ s,s+1\right] }=\left\{ R|\sqrt{2\sigma \left(
P_{s},R\right) }+\sqrt{2\sigma \left( R,P_{s+1}\right) }=\sqrt{2\sigma
\left( P_{s},P_{s+1}\right) }\right\}  \label{a2.6}
\end{equation}

The links $\mathbf{P}_{s}\mathbf{P}_{s+1}$ of the world chain have the same
length. According to (\ref{a2.1}) it means that they have the same mass. If
the particle motion is free, the adjacent vectors $\mathbf{P}_{s}\mathbf{P}%
_{s+1}$ and $\mathbf{P}_{s+1}\mathbf{P}_{s+2}$ are in parallel. It means
that 
\begin{equation}
\left( \mathbf{P}_{s}\mathbf{P}_{s+1}.\mathbf{P}_{s+1}\mathbf{P}%
_{s+2}\right) =\left\vert \mathbf{P}_{s}\mathbf{P}_{s+1}\right\vert \cdot
\left\vert \mathbf{P}_{s+1}\mathbf{P}_{s+2}\right\vert ,\qquad s=0,\pm 1,\pm
2,...  \label{a2.7}
\end{equation}%
where $\left( \mathbf{P}_{s}\mathbf{P}_{s+1}.\mathbf{P}_{s+1}\mathbf{P}%
_{s+2}\right) $ is the scalar product of vectors $\mathbf{P}_{s}\mathbf{P}%
_{s+1}$ and $\mathbf{P}_{s+1}\mathbf{P}_{s+2}$.

The scalar product $\left( \mathbf{PQ}.\mathbf{RS}\right) $ of two vectors $%
\mathbf{PQ}$ and $\mathbf{RS}$ in physical geometry is defined by the
relation%
\begin{equation}
\left( \mathbf{PQ}.\mathbf{RS}\right) =\sigma \left( P,S\right) +\sigma
\left( Q,R\right) -\sigma \left( P,R\right) -\sigma \left( Q,S\right)
\label{a2.8}
\end{equation}%
where $P,Q,R,S$ are the points, which determine the vectors $\mathbf{PQ}$
and $\mathbf{RS}$\textbf{.} In the $\ $proper Euclidean geometry the
definition of the scalar product (\ref{a2.8}) is equivalent to the
conventional definition of the scalar product in the linear vector space.
The definition (\ref{a2.8}) via world function does not refer to the linear
vector space, and it may be used in the case of such a physical geometry,
where one cannot introduce a linear vector space.

Equivalence (parallelism and equality of lengths) of two vectors $\mathbf{P}%
_{s}\mathbf{P}_{s+1}$ and $\mathbf{P}_{s+1}\mathbf{P}_{s+2}$ is written in
the form of two equations 
\begin{equation}
\left( \mathbf{P}_{s}\mathbf{P}_{s+1}.\mathbf{P}_{s+1}\mathbf{P}%
_{s+2}\right) \equiv \sigma \left( P_{s},P_{s+2}\right) -\sigma \left(
P_{s},P_{s+1}\right) -\sigma \left( P_{s+1},P_{s+2}\right) =2\sigma \left(
P_{s},P_{s+1}\right)  \label{a2.9}
\end{equation}%
\begin{equation}
\sigma \left( P_{s},P_{s+1}\right) =\sigma \left( P_{s+1},P_{s+2}\right)
,\qquad s=0,\pm 1,\pm 2,...  \label{a2.10}
\end{equation}%
By means of relation (\ref{a2.10}) the equation (\ref{a2.9}) can be reduced
to the form%
\begin{equation}
\sigma \left( P_{s},P_{s+2}\right) =4\sigma \left( P_{s},P_{s+1}\right)
,\qquad s=0,\pm 1,\pm 2,...  \label{a2.11}
\end{equation}

Two equations (\ref{a2.10}), (\ref{a2.11}) describe the world chain of a
free microparticle. In the space-time of Minkowski the timelike links (\ref%
{a2.6}) of this chain are segments of the straight line. These segments
became infinitesimal, if the lengths $\left\vert \mathbf{P}_{s}\mathbf{P}%
_{s+1}\right\vert $ of links tend to zero, and world chain transforms into a
world line in space-time of Minkowski. At such a limit the geometrical
length $\mu \rightarrow 0$, and the relation (\ref{a2.2}) cannot be used for
geometrization of the finite mass $m$ of the particle. In this case the
particle mass $m$ becomes some external characteristic of a particle, which
is not connected with the space-time geometry directly.

In the case of arbitrary space-time geometry the link (\ref{a2.6}) is not a
segment of one-dimensional straight, in general. Indeed, according to
definition (\ref{a2.6}) the link $\mathcal{T}_{\left[ P_{s}P_{s+1}\right] }$
is a three-dimensional surface in the 4-dimensional space-time. In the case
of the space-time of Minkowski and timelike vector $\mathbf{P}_{s}\mathbf{P}%
_{s+1}$ the surface $\mathcal{T}_{\left[ P_{s}P_{s+1}\right] }$ degenerates
into a segment of one-dimensional straight line. The same degeneration takes
place in the case of the Riemannian space-time geometry.

This degeneration is connected with original suppositions on the space-time
geometry and on geometry, in general. One supposes, that the geometry is
single-variant and infinitely divisible. In reality the space-time geometry
is multivariant and restrictedly divisible.

Let us explain concept of multivariance. Two vectors $\mathbf{P}_{0}\mathbf{P%
}_{1}$ and $\mathbf{Q}_{0}\mathbf{Q}_{1}$ are equivalent (equal) ($\mathbf{P}%
_{0}\mathbf{P}_{1}$eqv$\mathbf{Q}_{0}\mathbf{Q}_{1}$), if they are in
parallel and their length are equal. 
\begin{equation}
\left( \mathbf{P}_{0}\mathbf{P}_{1}.\mathbf{Q}_{0}\mathbf{Q}_{1}\right)
=\left\vert \mathbf{P}_{0}\mathbf{P}_{1}\right\vert \cdot \left\vert \mathbf{%
Q}_{0}\mathbf{Q}_{1}\right\vert  \label{a2.12}
\end{equation}%
\begin{equation}
\left\vert \mathbf{P}_{0}\mathbf{P}_{1}\right\vert =\left\vert \mathbf{Q}_{0}%
\mathbf{Q}_{1}\right\vert  \label{a2.14}
\end{equation}%
In terms of the world function the relations (\ref{a2.12}), (\ref{a2.14})
are written in the form%
\begin{equation}
\sigma \left( P_{0},Q_{1}\right) +\sigma \left( P_{1},Q_{0}\right) -\sigma
\left( P_{0},Q_{0}\right) -\sigma \left( P_{1},Q_{0}\right) =2\sigma \left(
P_{0},P_{1}\right)  \label{a2.15}
\end{equation}%
\begin{equation}
\sigma \left( P_{0},P_{1}\right) =\sigma \left( Q_{0},Q_{1}\right)
\label{a2.16}
\end{equation}%
The definition of equivalence of two vectors does not refer to a coordinate
system, or to dimension of the space-time. If vector $\mathbf{P}_{0}\mathbf{P%
}_{1}$ is given, and we are going to determine vector $\mathbf{Q}_{0}\mathbf{%
Q}_{1}$ at the point $Q_{0}$, which is equivalent to vector $\mathbf{P}_{0}%
\mathbf{P}_{1}$, we should solve the system of two equations (\ref{a2.15}), (%
\ref{a2.16}) with respect to the point $Q_{1}$ at given points $%
Q_{0},P_{0},P_{1}$.

In the proper Euclidean space and in the space of Minkowski for timelike
vector $\mathbf{P}_{0}\mathbf{P}_{1}$ ($\sigma \left( P_{0},P_{1}\right) >0$%
) one obtains one and only one solution $Q_{1}$ of the two equations (\ref%
{a2.15}), (\ref{a2.16}). This fact is formulated as follows. The geometry of
Minkowski is single-variant with respect to any point $Q_{0}$ and with
respect to any timelike vector $\mathbf{P}_{0}\mathbf{P}_{1}$.

However, the same geometry of Minkowski is multivariant with respect to any
point $Q_{0}$ and any spacelike vector $\mathbf{P}_{0}\mathbf{P}_{1}$ ($%
\sigma \left( P_{0},P_{1}\right) <0$). It means, that in the case of given
point $Q_{0}$ and given spacelike vector $\mathbf{P}_{0}\mathbf{P}_{1}$ the
system of two equations (\ref{a2.15}), (\ref{a2.16}) has many solutions for
the point $Q_{1}$.

All axiomatizable geometries, constructed on the basis of the linear vector
space appear to be single-variant, because the construction of the linear
vector space does not admit multivariance. The fact is that, in a
multivariant geometry the equivalence relation is intransitive, whereas
axioms of the linear vector space demands a transitive equivalence relation.
The axiomatizable geometries cannot be multivariant, because they are
constructed on the basis of the linear vector space. Thus, multivariance and
intransitivity of the equivalence relation are incompatible with
axiomatizability of a geometry.

In the nonaxiomatizable space-time geometry the links $\mathcal{T}_{\left[
P_{s}P_{s+1}\right] }$ of the world chain (\ref{a2.5}) are surfaces. If the
space-time geometry is close to the geometry of Minkowski, the links $%
\mathcal{T}_{\left[ P_{s}P_{s+1}\right] }$ have the shape of narrow tubes.
If the space-time geometry tends to the geometry of Minkowski, these tubes
degenerate into segments of one-dimensional straight line. However, such a
degeneration takes place only, if vectors $\mathbf{P}_{s}\mathbf{P}_{s+1}$
are timelike.

In the case of spacelike vectors $\mathbf{P}_{s}\mathbf{P}_{s+1}$ any link $%
\mathcal{T}_{\left[ P_{s}P_{s+1}\right] }$ has the shape of infinite
3-dimensional hyperplanes, which are tangent to the light cone and contain
the points of one-dimensional segment, restricted by points $P_{s},P_{s+1}$.
Such a shape of spacelike links $\mathcal{T}_{\left[ P_{s}P_{s+1}\right] }$
is a reason, why there are only timelike world lines. Spacelike world lines
have not been discovered. At the conventional approach to space-time
geometry absence of spacelike world lines is simply postulated.

It is worth to note, that the multivariant space-time geometries exist
indeed. In such a space-time geometry world chains appear to be stochastic,
and one needs to use statistical description to obtain a deterministic
description and to make some prediction on possible evolution of a particle.

For instance, the world function%
\begin{equation}
\sigma _{\mathrm{d}}=\sigma _{\mathrm{M}}+d\cdot \text{sign}\sigma _{\mathrm{%
M}},\qquad d=\frac{\hbar }{2bc}=\text{const}  \label{a2.17}
\end{equation}%
where $\sigma _{\mathrm{M}}$ is the world function (\ref{a2.3}) of the
Minkowski geometry, describes the space-time geometry, which is multivariant
with respect to timelike vectors. Here $\hbar $ is the quantum constant and $%
b$ is the universal constant, defined by the relation (\ref{a2.2}). World
chains are stochastic in this space-time geometry. Statistical description
of stochastic timelike world chains is equivalent to the quantum description
in terms of the Schr\"{o}dinger equation \cite{R1991}. The quantum constant $%
\hbar $ appears in the description from the expression (\ref{a2.17}) for the
world function, whereas the universal constant $b$ disappears, because the
statistical description is sensitive to the length $\mu $ of links of the
world chain. Replacement of $\mu $ by its expression $\mu =m/b$, which
follows form (\ref{a2.2}), leads to disappearance of $b$ and appearance of $%
m $ instead of $\mu $.

It should note, that the geometry (\ref{a2.17}) is uniform, isotropic and
discrete. At conventional approach to a geometry as a logical construction,
an isotropic geometry cannot be discrete. Such a viewpoint takes place,
because the conventional approach connects any discreteness with properties
of the manifold. According to this approach a discrete geometry cannot be
given on continual manifold. In reality, a discreteness is determined by
properties of the world function. On the same manifold of Minkowski one can
define both a continual geometry and a discrete one. The space-time geometry
(\ref{a2.17}) is discrete, because in this geometry there are no vectors $%
\mathbf{PQ}$ of the length $\left\vert \mathbf{PQ}\right\vert $, satisfying
the condition 
\begin{equation}
0<\left\vert \mathbf{PQ}\right\vert ^{2}<d^{2}  \label{a2.18}
\end{equation}%
where $d$ is the constant, defined in (\ref{a2.17}). Of course, such a
space-time should be qualified as a discrete. See details in \cite{R2008}.

\section{General properties of dynamic equations}

Dynamic equations (\ref{a2.10}), (\ref{a2.11}) describe evolution of
microparticle state, which is described by the vector $\mathbf{P}_{s}\mathbf{%
P}_{s+1}$. The dynamic equations are finite difference equations (not
differential). They are sensitive to the length of a step (link of the
chain). They are written in the form, which is insensitive to a choice of a
coordinate system and to the dimension of the space-time. All information on
dynamic equations is concentrated in the space-time geometry and in the
length of links of the world chain. We are going to rewrite equations (\ref%
{a2.10}), (\ref{a2.11}) in the form of differential equations, tending the
length of links to zero. We are interested in the form of dynamic equations
in the case of non-Riemannian space-time geometry. In particular, we are
interested in the form of dynamic equations in the case, when the space-time
geometry is deformed by a presence of a heavy sphere.

Before to write dynamic equations (\ref{a2.10}), (\ref{a2.11}) in the form
of differential equations, we rewrite them in terms of distance $\rho =\sqrt{%
2\sigma }$. We obtain%
\begin{eqnarray}
\rho \left( P_{s},P_{s+1}\right) &=&\rho \left( P_{s+1},P_{s+2}\right)
,\qquad s=0,\pm 1,\pm 2,...  \label{a2.19} \\
\rho \left( P_{s},P_{s+2}\right) &=&2\rho \left( P_{s},P_{s+1}\right)
,\qquad s=0,\pm 1,\pm 2,..  \label{a2.20}
\end{eqnarray}%
In the case of the proper Euclidean space all points $P_{s}$, $s=0,\pm 1,\pm
2,...$ lie on one straight line. In the space-time geometry of Minkowski all
points $P_{s}$, $s=0,\pm 1,\pm 2,...$ lie on one timelike straight, provided 
$\rho ^{2}\left( P_{s},P_{s+1}\right) =\rho ^{2}\left( P_{0},P_{1}\right) >0$%
.

The equations (\ref{a2.19}), (\ref{a2.20}), as well equations (\ref{a2.10}),
(\ref{a2.11}) realize the procedure of the straight line construction by
means of only compasses. One starts from the segment $P_{0}P_{1}$ of length $%
\rho _{0}$ and draws a sphere $\mathcal{S}_{P_{1,}\rho _{0}}$ of radius $%
\rho _{0}$ with the center at the point $P_{1}$. Besides, one draws a sphere 
$\mathcal{S}_{P_{0},2\rho _{0}}$ of radius $2\rho _{0}$ with the center at
the point $P_{0}$. The spheres $\mathcal{S}_{P_{1},\rho _{0}}$ and $\mathcal{%
S}_{P_{0},2\rho _{0}}$ have the only common point $P_{2}$. It is the point $%
P_{2}$, which is defined as a common point of spheres $\mathcal{S}%
_{P_{1},\rho _{0}}$ and $\mathcal{S}_{P_{0},2\rho _{0}}$, which are tangent
to each other at this point. The segment $P_{1}P_{2}$ has the length $\rho
_{0}$. One draws spheres $\mathcal{S}_{P_{2},\rho _{0}}$ and $\mathcal{S}%
_{P_{1},2\rho _{0}}$, which has the only common point $P_{3}$ and so on. All
points , constructed by this method, lie on the same straight line. This
procedure is sensitive to an error of the radius $\rho _{0}$ in the sense,
that the error $\delta \rho _{0}=\alpha \rho _{0}$, $0<\alpha \ll 1$ of the
radius generates the error $\delta \rho $ of the point $P_{2}$ position,
which is larger, than $\delta \rho _{0}$ ( $\delta \rho =\sqrt{\alpha }\rho
_{0}\gg \delta \rho _{0}$).

One can construct points $P_{0},P_{1},P_{2},...$ which lie on the same
straight in the proper Euclidean space, if one uses the fact, that adjacent
vectors $\mathbf{P}_{s}\mathbf{P}_{s+1}$ and $\mathbf{P}_{s+1}\mathbf{P}%
_{s+2}$ are in parallel. However, in this case one uses usually parallelism,
defined in the linear vector space. This definition of parallelism refers to
a coordinate system and to the dimension of the Euclidean space. This fact
prevents from a use of dynamic equations in the case of arbitrary space-time
geometry, where one cannot introduce a linear vector space. Dynamic equation
(\ref{a2.20}) has been obtained from the equation (\ref{a2.9}), which
describes parallelism of vectors $\mathbf{P}_{s}\mathbf{P}_{s+1}$ and $%
\mathbf{P}_{s+1}\mathbf{P}_{s+2}$. However, in this case one uses the
definition of parallelism in the form (\ref{a2.7}), which refers to the
world function only. As a result the form of dynamic equations (\ref{a2.10}%
), (\ref{a2.11}) appears to be valid in the case of arbitrary physical
space-time geometry.

Let us consider dynamic equations (\ref{a2.10}), (\ref{a2.11}) in the
space-time, whose geometry is close to that of Minkowski. For simplicity we
consider the case of the particle at rest, where coordinates of the points $%
P_{0},P_{1}$ are 
\begin{equation}
P_{0}=\left\{ -\mu ,\mathbf{0}\right\} ,\qquad P_{1}=\left\{ 0,\mathbf{0}%
\right\} ,,\qquad P_{2}=\left\{ t,\mathbf{x}\right\}  \label{a2.21}
\end{equation}%
Equations of "spheres" $\mathcal{S}_{P_{0},2\mu }$ and $\mathcal{S}%
_{P_{1},\mu }$ are 
\begin{equation}
\mathcal{S}_{P_{1},\mu }:t^{2}-\mathbf{x}^{2}+\alpha _{1}\left( t,\mathbf{x}%
\right) \mu ^{2}=\mu ^{2},\qquad \left\vert \alpha _{1}\left( t,\mathbf{x}%
\right) \right\vert \ll 1  \label{a2.22}
\end{equation}%
\begin{equation}
\mathcal{S}_{P_{0},2\mu }:\left( t+\mu \right) ^{2}-\mathbf{x}^{2}+\alpha
_{2}\left( t,\mathbf{x}\right) \mu ^{2}=\left( 2\mu \right) ^{2},\qquad
\left\vert \alpha _{2}\left( t,\mathbf{x}\right) \right\vert \ll 1
\label{a2.23}
\end{equation}%
where the small quantities $\alpha _{1}\left( t,\mathbf{x}\right) \mu ^{2}$
and $\alpha _{2}\left( t,\mathbf{x}\right) \mu ^{2}$ take into account a
small deflection of the space-time geometry from the geometry of Minkowski.
In fact the "spheres" $\mathcal{S}_{P_{0},2\mu }$ and $\mathcal{S}%
_{P_{1},\mu }$ are deformed spheres, i.e. the surfaces, which are close to
spheres.

The time coordinate $t$ of the intersection point is defined by the relation%
\begin{equation}
\left( t+\mu \right) ^{2}-t^{2}=3\mu ^{2}-\left( \alpha _{2}-\alpha
_{1}\right) \mu ^{2}  \label{a2.24}
\end{equation}%
or 
\begin{equation}
t=\mu \left( 1-\frac{\alpha _{2}\left( t,\mathbf{x}\right) -\alpha
_{1}\left( t,\mathbf{x}\right) }{2}\right)  \label{a2.25}
\end{equation}%
In the space-time of Minkowski, where $\alpha _{1}=\alpha _{2}=0$, one
obtains $t=\mu ,\ \mathbf{x=0}$. It means that three points $%
P_{0},P_{1},P_{2}$ lie on the same timelike straight line.

One obtains from (\ref{a2.25}) and (\ref{a2.22}), that the spatial
coordinates $\mathbf{x}$ of the intersection point are placed on the
two-dimensional surface 
\begin{equation}
\mathbf{x}^{2}=\left( -\alpha _{2}+\frac{\left( \alpha _{2}-\alpha
_{1}\right) ^{2}}{4}\right) \mu ^{2}  \label{a2.26}
\end{equation}

If%
\begin{equation}
\alpha _{2}<0\text{\ \ and\ \ }\left\vert \alpha _{2}\right\vert ,\left\vert
\alpha _{1}\right\vert \ll 1  \label{a2.27}
\end{equation}%
there are many intersection points, placed in the small spatial region with
radius of the order $\sqrt{\left\vert \alpha _{2}\right\vert }\mu .$ In this
case the world chain of the particle is multivariant (stochastic).

In the case, when 
\begin{equation}
\alpha _{2}>\frac{\left( \alpha _{2}-\alpha _{1}\right) ^{2}}{4}
\label{a2.28}
\end{equation}%
there are no intersection points between the spheres $\mathcal{S}%
_{P_{0},2\mu }$ and $\mathcal{S}_{P_{1},\mu }$, defined by the relations (%
\ref{a2.23}), (\ref{a2.22}). It means, that the world chain with geometrical
mass $\mu $ (length of the chain link) cannot exist. Such a region of the
space-time is a "dead region" for microparticles of the geometrical mass $%
\mu $.

The most interesting case is realized, when rhs of (\ref{a2.26}) vanishes.
In this case the links of the world chain are determined uniquely. The world
chain appears to be deterministic. Deterministic (single-variant) world
chain appears in the space-time geometry of Minkowski and in the Riemannian
space-time geometry.

The result on stochasticity of world chains, which are obtained in this
section, are valid only for a simple microparticle (elementary particle),
which is described by the skeleton $\mathcal{P}^{1}=\left\{
P_{0},P_{1}\right\} $, consisting of two points. Macroparticles (metorites,
planets, stars) consist of many microparticles, connected between themselves
by some force fields. Microparticles cannot move independently and
stochastically, because they are connected between themselves. As a result
the motion of all microparticles inside the macroparticle is not free. It is
described by a deterministic (single-valued) world chain. Any such a
deterministic world chain is determined as a mean chain, which is a result
of averaging over the surface (\ref{a2.26}). A result of this averaging of
the surface (\ref{a2.26}) leads to the point $P_{2}=\left\{ t,\mathbf{x}%
\right\} $, which together with the given point $P_{1}=\left\{ 0,\mathbf{0}%
\right\} $ determines uniquely the next link $\left( P_{1},P_{2}\right) $.

Thus, to obtain a deterministic world chain of a macroparticle, one needs to
produce an averaging of stochastic world chains of microparticles, which is
equivalent to a statistical description. The obtained mean world chains are
described by dynamic equations in finite difference, which depend on the
geometrical mass $\mu $ of microparticles, constituting the macroparticle.
To obtain differential dynamic equations, one needs to go to the limit $\mu
\rightarrow 0$ in finite difference equations of dynamics. As a result one
obtains differential dynamic equations, describing free motion of
macroparticles in the given physical space-time.

Multivariant physical space-time geometries are not considered in the
contemporary theoretical physics. One believes, that the space-time geometry
may not be nonaxiomatizable (multivariant), because it is not known, how to
construct such geometries (the deformation principle is either unknown, or
is not accepted). One does not admit, that a free motion of microparticles
may be stochastic, and statistical averaging of these stochastic world
chains is not considered. One believes, that the mean world lines of free
macroparticles are geodesic lines of the space-time geometry. Restoring the
space-time geometry on the basis of these geodesic lines one obtains the
Riemannian geometry, which is determined by its geodesic lines and lengths
of their segments. One ignores the fact, that there is an intermediate link
between the world lines of free macroparticles and the space-time geometry.
This intermediate link has a form of statistical averaging. As a result
different physical geometries of space-time, having similar mean world lines
of free macroparticles, are substituted by one (Riemannian) geometry.

Such an approach is admissible, when one considers motion of free
macroparticles in a fixed space-time geometry. However, this approach appear
to be wrong, when a generation of the space-time geometry by the matter
distribution is considered. For instance, the world function of the
space-time geometry, generated by a heavy sphere gives in the first
approximation \cite{R2009} 
\begin{equation}
\sigma _{\left( 1\right) }\left( t_{1},\mathbf{x}_{1};t_{2},\mathbf{x}%
_{2}\right) =\frac{1}{2}\left( c^{2}\left( 1-\frac{4GM}{c^{2}\left\vert 
\mathbf{x}_{2}\mathbf{+x}_{1}\right\vert }\right) \left( t_{2}-t_{1}\right)
^{2}-\left( \mathbf{x}_{2}\mathbf{-x}_{1}\right) ^{2}\right)   \label{a2.29}
\end{equation}%
In the second approximation we obtain \cite{R2009}%
\begin{equation}
\sigma _{\left( 2\right) }\left( t_{1},\mathbf{x}_{1};t_{2},\mathbf{x}%
_{2}\right) =\frac{1}{2}\left( c^{2}\left( t_{2}-t_{1}\right) ^{2}-\left( 
\mathbf{x}_{2}-\mathbf{x}_{1}\right) ^{2}\right) +\delta \sigma _{2}\left(
t_{1},\mathbf{x}_{1};t_{2},\mathbf{x}_{2}\right)   \label{a2.30}
\end{equation}%
where%
\begin{equation}
\delta \sigma _{2}\left( t_{1},\mathbf{x}_{1};t_{2},\mathbf{x}_{2}\right)
=V_{2}\left( \mathbf{x}_{1},\mathbf{x}_{2}\right) \left( c\left(
t_{1}-t_{2}\right) +A_{2}\left( \mathbf{x}_{1},\mathbf{x}_{2}\right)
\left\vert \frac{\mathbf{x}_{1}+\mathbf{x}_{2}}{2}\right\vert \right) ^{2}
\label{a2.30a}
\end{equation}%
\begin{equation}
V_{2}\left( \mathbf{x}_{1},\mathbf{x}_{2}\right) =-4\frac{GM}{%
c^{2}\left\vert \mathbf{x}_{1}+\mathbf{x}_{2}\right\vert }\frac{\sqrt{\left(
1-\frac{8GM}{c^{2}\left\vert \mathbf{x}+\mathbf{x}_{2}\right\vert }\right) }%
\left( 1-\frac{4GM}{c^{2}\left\vert \mathbf{x}_{1}\right\vert }\right)
^{2}\left( 1-\frac{4GM}{c^{2}\left\vert \mathbf{x}_{2}\right\vert }\right)
^{2}}{\left( 1-c^{-2}\phi \left( 0\right) \right) \left( 1-\frac{4GM}{%
c^{2}\left\vert \mathbf{x}_{1}+\mathbf{x}_{2}\right\vert }\right) \left( 1-%
\frac{2GM}{c^{2}\left\vert \mathbf{x}_{1}\right\vert }-\frac{2GM}{%
c^{2}\left\vert \mathbf{x}_{2}\right\vert }\right) ^{2}}  \label{a2.31}
\end{equation}%
\begin{equation}
A_{2}\left( \mathbf{x}_{1},\mathbf{x}_{2}\right) =\frac{\frac{4GM}{%
c^{2}\left\vert \mathbf{x}_{2}\right\vert }-\frac{4GM}{c^{2}\left\vert 
\mathbf{x}_{1}\right\vert }}{\left( 1-\frac{4GM}{c^{2}\left\vert \mathbf{x}%
_{1}\right\vert }\right) \left( 1-\frac{4GM}{c^{2}\left\vert \mathbf{x}%
_{2}\right\vert }\right) }\sqrt{1-\frac{8GM}{c^{2}\left\vert \mathbf{x}_{1}+%
\mathbf{x}_{2}\right\vert }}  \label{a2.31a}
\end{equation}%
and $\phi \left( 0\right) $ is the Newtonian gravitational potential in the
center of the heavy sphere of the mass $M$, which is placed at the origin of
the coordinate system.

Comparison of the world function (\ref{a2.29}) of the first approximation,
with the world function (\ref{a2.30}) of the second approximation (\ref%
{a2.30}) - (\ref{a2.31a}) shows that the iteration process is to converge
rapidly, if the gravitational potential is slight, i.e. \label{0end}%
\begin{equation}
\frac{2GM}{c^{2}\left\vert \mathbf{x}\right\vert }\ll 1  \label{a2.33}
\end{equation}

The well known Schwarzchild solution for gravitational field of a heavy
sphere has the form%
\begin{equation}
ds^{2}=\left( 1-\frac{2GM}{c^{2}r}\right) c^{2}dt^{2}-\left( 1-\frac{2GM}{%
c^{2}r}\right) ^{-1}dr^{2}-r^{2}\left( d\theta ^{2}+\sin ^{2}\theta d\varphi
^{2}\right)  \label{a2.34}
\end{equation}
Gravitational potential $V\left( r\right) =\frac{2GM}{r}$ of (\ref{a2.34})
coincides with the gravitational potential of the first approximation (\ref%
{a2.29})%
\begin{equation}
V_{1}\left( \mathbf{x}_{1},\mathbf{x}_{2}\right) =\frac{4GM}{\left\vert 
\mathbf{x}_{1}+\mathbf{x}_{2}\right\vert }  \label{a2.32}
\end{equation}%
However it distinguishes from the gravitational potential (\ref{a2.31}) of
the second approximation.

Comparison of expressions (\ref{a2.29}) - (\ref{a2.31}) with exact
Schwarzchild solution (\ref{a2.34}) shows the following differences.

1. Geometry (\ref{a2.29}) - (\ref{a2.31}) is not a Riemannian geometry.
Corresponding world function is obtained directly, whereas in the case of
the Schwarzchild solution (\ref{a2.34}) the space-time geometry is \textit{%
supposed to be Riemannian. }Under this supposition it is obtained by means
of the metric tensor, which is obtained as a solution of the gravitational
equations.

2. Potential (\ref{a2.31}) in the metric component $g_{00}$ depends on the
particle mass $M$ linearly in the Schwarzchild solution, whereas this
dependence is not linear in the case of geometry (\ref{a2.30}) - (\ref%
{a2.31a}).

Both equations (\ref{a2.34}) and (\ref{a2.30}) - (\ref{a2.31a}) cannot be
true. Although the world function (\ref{a2.30}) - (\ref{a2.31}) is only a
second approximation, (but not an exact solution), it is closer to the
truth, than the Schwarzchild solution (\ref{a2.34}), because the
Schwarzchild solution is based on a use of inconsistent Riemannian geometry.

In the light of hesitations in consistency of the Riemannian geometry the
conclusion on existence of the dark matter and other astrophysical
conclusions, based on the contemporary (Riemannian) theory of gravitation
may appear to be a little too previous.

\section{Dynamic equations for free particle in the space-time of Minkowski}

At first, we consider application of suggested method to the case of
space-time geometry of Minkowski. This method transforms dynamic equations,
written in terms of finite differences to differential equations of
dynamics. Although the obtained result is trivial, it is interesting in the
sense, that it forbids an existence of spacelike world lines.

We consider two connected links of the world chain, defined by the points $%
P_{0},P_{1},P_{2},$having coordinates%
\begin{equation}
P_{0}=\left\{ y-dy_{1}\right\} ,\qquad P_{1}=\left\{ y\right\} ,\qquad
P_{2}=\left\{ y+dy_{2}\right\}  \label{a3.1}
\end{equation}%
where 
\begin{equation}
y=\left\{ t,\mathbf{y}\right\} ,\qquad dy_{1}=\left\{ dt_{1},d\mathbf{y}%
_{1}\right\} ,\qquad dy_{2}=\left\{ dt_{2},d\mathbf{y}_{2}\right\}
\label{a3.2}
\end{equation}%
are coordinates in some inertial coordinate system, where the world function
has the form (\ref{a2.3}). Dynamic equations (\ref{a2.10}), (\ref{a2.11})
have the form%
\begin{eqnarray}
\mathrm{\ }\sigma _{\mathrm{M}}\left( y,y-dy_{1}\right) &=&\sigma _{\mathrm{M%
}}\left( y,y+dy_{2}\right)  \label{a3.3} \\
\mathrm{\ }4\sigma _{\mathrm{M}}\left( y,y-dy_{1}\right) &=&\sigma _{\mathrm{%
M}}\left( y-dy_{1},y+dy_{2}\right)  \label{a3.4}
\end{eqnarray}%
In the developed form one obtains%
\begin{eqnarray}
\frac{1}{2}c^{2}\left( dt_{1}\right) ^{2}-\frac{1}{2}\left( d\mathbf{y}%
_{1}\right) ^{2} &=&\frac{1}{2}c^{2}\left( dt_{2}\right) ^{2}-\frac{1}{2}%
\left( d\mathbf{y}_{2}\right) ^{2}  \label{a3.5} \\
2c^{2}\left( dt_{1}\right) ^{2}-2\left( d\mathbf{y}_{1}\right) ^{2} &=&\frac{%
1}{2}c^{2}\left( dt_{1}+dt_{2}\right) ^{2}-\frac{1}{2}\left( d\mathbf{y}%
_{1}+d\mathbf{y}_{2}\right) ^{2}  \label{a3.6}
\end{eqnarray}

We introduce designations 
\begin{eqnarray}
\mathbf{v}_{1} &=&\frac{d\mathbf{y}_{1}}{dt_{1}},\qquad \mathbf{v}_{2}=\frac{%
d\mathbf{y}_{2}}{dt_{2}},\qquad \mathbf{\beta }_{1}\mathbf{=}\frac{\mathbf{v}%
_{1}}{c},\qquad \mathbf{\beta }_{2}\mathbf{=}\frac{\mathbf{v}_{2}}{c}
\label{a3.7} \\
\mathbf{\beta }_{1} &=&\mathbf{\beta -}\frac{1}{2}\mathbf{\dot{\beta}}%
dt,\qquad \mathbf{\beta }_{1}=\mathbf{\beta +}\frac{1}{2}\mathbf{\dot{\beta}}%
dt,\qquad \mathbf{\dot{\beta}\equiv }\frac{d\mathbf{\beta }}{dt},\qquad dt=%
\frac{dt_{1}+dt_{2}}{2}  \label{a3.8}
\end{eqnarray}%
where 
\begin{equation}
\mathbf{v=}c\mathbf{\beta }\qquad \mathbf{\dot{v}}=c\mathbf{\dot{\beta}}
\label{a3.8a}
\end{equation}
are the mean velocity and the mean acceleration of the particle on the
interval $\left( P_{0},P_{2}\right) $.

One rewrites equations (\ref{a3.5}), (\ref{a3.6}) in the form%
\begin{eqnarray}
1-\mathbf{\beta }_{1}^{2} &=&\frac{dt_{2}^{2}}{dt_{1}^{2}}-\mathbf{\beta }%
_{2}^{2}\frac{dt_{2}^{2}}{dt_{1}^{2}}  \label{a3.9} \\
\mathrm{\ }4-4\mathbf{\beta }_{1}^{2} &=&\left( 1+\frac{dt_{2}}{dt_{1}}%
\right) ^{2}-\left( \mathbf{\beta }_{1}+\mathbf{\beta }_{2}\frac{dt_{2}}{%
dt_{1}}\right) ^{2}  \label{a3.10}
\end{eqnarray}%
One obtains from equation (\ref{a3.9}) to within $\left( dt\right) ^{2}$%
\begin{equation}
\frac{dt_{2}^{2}}{dt_{1}^{2}}=\frac{1-\mathbf{\beta }_{1}^{2}}{1-\mathbf{%
\beta }_{2}^{2}}=\frac{1-\left( \mathbf{\beta -}\frac{1}{2}\mathbf{\dot{\beta%
}}dt\right) ^{2}}{1-\left( \mathbf{\beta +}\frac{1}{2}\mathbf{\dot{\beta}}%
dt\right) ^{2}}=1+2\frac{\mathbf{\beta \dot{\beta}}dt}{1-\mathbf{\beta }^{2}}%
+\frac{2\left( \mathbf{\beta \dot{\beta}}dt\right) ^{2}}{\left( 1-\mathbf{%
\beta }^{2}\right) ^{2}}=1+\alpha  \label{a3.11}
\end{equation}%
where%
\begin{equation}
\alpha =2\frac{\mathbf{\beta \dot{\beta}}dt}{1-\mathbf{\beta }^{2}}+2\frac{%
\left( \mathbf{\beta \dot{\beta}}dt\right) ^{2}}{\left( 1-\mathbf{\beta }%
^{2}\right) ^{2}}+O\left( dt^{3}\right)  \label{a3.12}
\end{equation}%
Then%
\begin{equation}
\frac{dt_{2}}{dt_{1}}=1+\frac{1}{2}\alpha -\frac{1}{8}\alpha ^{2}
\label{a3.14}
\end{equation}%
Substituting (\ref{a3.8}) and (\ref{a3.14}) in (\ref{a3.10}), one obtains

\[
4\left( 1-\left( \mathbf{\beta -}\frac{1}{2}\mathbf{\dot{\beta}}dt\right)
^{2}\right) =\left( 2+\frac{\mathbf{\beta \dot{\beta}}dt}{1-\mathbf{\beta }%
^{2}}+\frac{1}{2}\frac{\left( \mathbf{\beta \dot{\beta}}dt\right) ^{2}}{%
\left( 1-\mathbf{\beta }^{2}\right) ^{2}}\right) ^{2} 
\]%
\begin{equation}
-\left( 2\mathbf{\beta }+\left( \mathbf{\beta +}\frac{1}{2}\mathbf{\dot{\beta%
}}dt\right) \left( 1+\frac{\mathbf{\beta \dot{\beta}}dt}{1-\mathbf{\beta }%
^{2}}+\frac{1}{2}\frac{\left( \mathbf{\beta \dot{\beta}}dt\right) ^{2}}{%
\left( 1-\mathbf{\beta }^{2}\right) ^{2}}\right) \right) ^{2}  \label{a3.15}
\end{equation}

After simplification one obtains%
\begin{equation}
\mathbf{-\dot{\beta}}^{2}dt^{2}=+3\frac{\left( \mathbf{\beta \dot{\beta}}%
dt\right) ^{2}}{\left( 1-\mathbf{\beta }^{2}\right) ^{2}}-3\mathbf{\beta }%
^{2}\left( \frac{\mathbf{\beta \dot{\beta}}dt}{1-\mathbf{\beta }^{2}}\right)
^{2}-2\frac{\left( \mathbf{\beta \dot{\beta}}dt\right) ^{2}}{1-\mathbf{\beta 
}^{2}}  \label{a3.16}
\end{equation}%
Note, that terms, which are proportional $dt$ disappear from the equation (%
\ref{a3.16}). After simplification the equation (\ref{a3.16}) takes the form 
\begin{equation}
\mathbf{\dot{\beta}}^{2}dt^{2}+\frac{\left( \mathbf{\beta \dot{\beta}}%
dt\right) ^{2}}{1-\mathbf{\beta }^{2}}=0  \label{a3.16a}
\end{equation}

Let us introduce designation%
\begin{equation}
\mathbf{\beta \dot{\beta}=}\sqrt{\mathbf{\beta }^{2}\mathbf{\dot{\beta}}^{2}}%
\cos \phi  \label{a3.18}
\end{equation}%
where $\phi $ is the angle between vectors $\mathbf{\beta }$ and $\mathbf{%
\dot{\beta}}$\textbf{.} The equation (\ref{a3.18}) takes the form 
\begin{equation}
\mathbf{\dot{\beta}}^{2}\left( 1+\frac{\mathbf{\beta }^{2}\cos ^{2}\phi }{1-%
\mathbf{\beta }^{2}}\right) =0  \label{a3.19}
\end{equation}

If the links of the world chain are timelike, then $\mathbf{\beta }^{2}=%
\mathbf{v}^{2}/c^{2}<1$, the expression in brackets of (\ref{a3.19}) is
positive, and the equation (\ref{a3.19}) can be satisfied only in the case,
when 
\begin{equation}
\mathbf{\dot{\beta}=}\frac{1}{c}\frac{d\mathbf{v}}{dt}=0,\qquad \mathbf{v}=%
\mathbf{v}\left( t\right) =\text{const}  \label{a3.20}
\end{equation}%
It is the expected unique solution. It should stress, that the unique
solutions obtained only in the case of timelike world chain. In the case of
spacelike world chain $\mathbf{\dot{\beta}}^{2}>1$, and there is such an
angle $\phi $ between vectors $\mathbf{\beta }$ and $\mathbf{\dot{\beta}}$,
that length $\left\vert \mathbf{\dot{\beta}}\right\vert $ of the vector $%
\mathbf{\dot{\beta}}$ is arbitrary. This angle is defined by the formula 
\begin{equation}
\cos ^{2}\phi =\frac{\mathbf{\beta }^{2}-1}{\mathbf{\beta }^{2}}<1
\label{a3.21}
\end{equation}%
There is no unique solution in the case of spacelike world chain.

Note, that for obtaining of differential equations from dynamic equations in
finite difference, we use the following representation of finite intervals $%
dy_{1}=\left\{ dt_{1}d\mathbf{y}_{1}\right\} $, $dy_{2}=\left\{ dt_{2}d%
\mathbf{y}_{2}\right\} $%
\[
d\mathbf{y}_{1}=c\mathbf{\beta }dt-\frac{c}{2}\mathbf{\dot{\beta}}\left(
dt\right) ^{2},\qquad d\mathbf{y}_{2}=c\mathbf{\beta }dt+\frac{c}{2}\mathbf{%
\dot{\beta}}\left( dt\right) ^{2} 
\]%
\[
dt=\frac{dt_{1}+dt_{2}}{2} 
\]

\section{Dynamic equations for motion of free particle in the gravitational
field of heavy sphere}

We consider the same equations (\ref{a3.3}), (\ref{a3.4}), but now in the
space-time geometry with the world function (\ref{a2.29})%
\begin{equation}
\sigma \left( t,\mathbf{y;}t^{\prime },\mathbf{y}^{\prime }\right) =\frac{1}{%
2}\left( c^{2}-\frac{2GM}{\sqrt{\mathbf{x}^{2}}}\right) \left( t-t^{\prime
}\right) ^{2}-\frac{1}{2}\left( \mathbf{y}-\mathbf{y}^{\prime }\right) ^{2}
\label{a4.1}
\end{equation}%
where 
\begin{equation}
\mathbf{x=}\frac{\mathbf{y+y}^{\prime }}{2}  \label{a4.2}
\end{equation}%
We shall use designations (\ref{a3.1}), (\ref{a3.2}) and (\ref{a3.7}), (\ref%
{a3.8}). Besides, we use the designations%
\begin{equation}
V=V\left( \mathbf{y}\right) =\frac{GM}{\sqrt{\left( \mathbf{y}\right) ^{2}}}%
,\qquad U=\frac{V}{c^{2}}  \label{a4.3}
\end{equation}

The equations (\ref{a3.3}), (\ref{a3.4}) with the world function (\ref{a4.1}%
) take the form 
\begin{equation}
\frac{1}{2}\left( 1-2U\left( \mathbf{y}-\frac{d\mathbf{y}_{1}}{2}\right)
\right) -\frac{1}{2}\mathbf{\beta }_{1}^{2}=\frac{1}{2}\left( 1-2U\left( 
\mathbf{y}+\frac{d\mathbf{y}_{2}}{2}\right) \right) \frac{dt_{2}^{2}}{%
dt_{1}^{2}}-\frac{1}{2}\mathbf{\beta }_{2}^{2}\frac{dt_{2}^{2}}{dt_{1}^{2}}
\label{a4.4}
\end{equation}%
\begin{eqnarray}
2\left( 1-2U\left( \mathbf{y}-\frac{d\mathbf{y}_{1}}{2}\right) \right) -2%
\mathbf{\beta }_{1}^{2} &=&\frac{1}{2}\left( 1-2U\left( \mathbf{y}+\frac{d%
\mathbf{y}_{2}-d\mathbf{\mathbf{y}_{1}}}{2}\right) \right) \left( 1+\frac{%
dt_{2}}{dt_{1}}\right) ^{2}  \nonumber \\
&&-\frac{1}{2}\left( \mathbf{\beta }_{1}+\mathbf{\beta }_{2}\frac{dt_{2}}{%
dt_{1}}\right) ^{2}  \label{a4.5}
\end{eqnarray}

It follows from (\ref{a4.4}), that%
\begin{equation}
\frac{dt_{2}^{2}}{dt_{1}^{2}}=\frac{1-2U\left( \mathbf{y}-\frac{d\mathbf{y}%
_{1}}{2}\right) -\mathbf{\beta }_{1}^{2}}{1-2U\left( \mathbf{y}+\frac{d%
\mathbf{y}_{2}}{2}\right) -\mathbf{\beta }_{2}^{2}}=1+\alpha +O\left(
dt^{3}\right)  \label{a4.6}
\end{equation}%
where $\alpha $ is an infinitesimal quantity. We shall consider $dt$ as a
principal infinitesimal quantity, and all infinitesimal quantities $d\mathbf{%
y}_{1},d\mathbf{y}_{2}$, $\alpha $ will be expressed via $\ dt$ and $\left(
dt\right) ^{2}$. The higher powers of $dt$ will be neglected.

One obtains from (\ref{a3.7}), (\ref{a3.8}) and (\ref{a4.6}) 
\begin{equation}
\frac{dt_{2}}{dt_{1}}=1+\frac{\alpha }{2}-\frac{\alpha ^{2}}{8},\qquad \frac{%
dt_{1}}{dt_{2}}=1-\frac{\alpha }{2}+\frac{3\alpha ^{2}}{8}  \label{a4.7}
\end{equation}%
\begin{equation}
\frac{dt}{dt_{1}}=\frac{1}{2}+\frac{1}{2}\frac{dt_{2}}{dt_{1}}=1+\frac{%
\alpha }{4}-\frac{\alpha ^{2}}{16}  \label{a4.8}
\end{equation}%
\begin{equation}
\frac{dt}{dt_{2}}=\frac{1}{2}+\frac{1}{2}\frac{dt_{1}}{dt_{2}}=1-\frac{%
\alpha }{4}+\frac{3}{16}\alpha ^{2}  \label{a4.9}
\end{equation}%
\begin{equation}
\frac{dt_{1}}{dt}=1-\frac{\alpha }{4}+\frac{\alpha ^{2}}{8},\qquad \frac{%
dt_{2}}{dt}=1+\frac{\alpha }{4}-\frac{\alpha ^{2}}{8}  \label{a4.10}
\end{equation}%
\begin{equation}
dt_{1}=\left( 1-\frac{\alpha }{4}+\frac{\alpha ^{2}}{8}\right) dt,\qquad
dt_{2}=\left( 1+\frac{\alpha }{4}-\frac{\alpha ^{2}}{8}\right) dt
\label{a4.11}
\end{equation}%
\begin{equation}
d\mathbf{y}_{1}=c\mathbf{\beta }_{1}dt_{1}=c\mathbf{\beta }_{1}\left( 1-%
\frac{\alpha }{4}\right) dt=c\left( \mathbf{\beta -}\frac{1}{2}\mathbf{\dot{%
\beta}}dt\right) \left( 1-\frac{\alpha }{4}\right) dt  \label{a4.12}
\end{equation}%
\begin{equation}
d\mathbf{y}_{2}=c\mathbf{\beta }_{2}dt_{2}=c\mathbf{\beta }_{2}\left( 1+%
\frac{\alpha }{4}\right) dt=c\left( \mathbf{\beta +}\frac{1}{2}\mathbf{\dot{%
\beta}}dt\right) \left( 1+\frac{\alpha }{4}\right) dt  \label{a4.14}
\end{equation}

Besides, the following decompositions are useful%
\begin{equation}
U\left( \mathbf{y}-\frac{d\mathbf{y}_{1}}{2}\right) =U\left( \mathbf{y}%
\right) +\delta _{1}U,\qquad U\left( \mathbf{y}+\frac{d\mathbf{y}_{2}}{2}%
\right) =U\left( \mathbf{y}\right) +\delta _{2}U  \label{a4.15}
\end{equation}%
\begin{equation}
U\left( \mathbf{y}+\frac{d\mathbf{y}_{2}-d\mathbf{\mathbf{y}_{1}}}{2}\right)
=U\left( \mathbf{y}\right) +\delta _{2-1}U  \label{a4.16}
\end{equation}%
where%
\begin{eqnarray}
\delta _{1}U &=&-\frac{d\mathbf{y}_{1}}{2}\mathbf{\nabla }U+\frac{1}{8}%
dy_{1}^{\alpha }dy_{1}^{\beta }U_{,\alpha \beta },\qquad U_{,\alpha \beta }=%
\frac{\partial ^{2}U\left( \mathbf{y}\right) }{\partial y^{\alpha }\partial
y^{\beta }}  \label{a4.17} \\
\delta _{2}U &=&\frac{d\mathbf{y}_{2}}{2}\mathbf{\nabla }U+\frac{1}{8}%
dy_{2}^{\alpha }dy_{2}^{\beta }U_{,\alpha \beta }  \label{a4.18}
\end{eqnarray}%
\begin{equation}
\delta _{2-1}U=\frac{d\mathbf{y}_{2}-d\mathbf{\mathbf{y}_{1}}}{2}\mathbf{%
\nabla }U+\frac{1}{8}\left( dy_{2}^{\alpha }-dy_{1}^{\alpha }\right) \left(
dy_{2}^{\beta }-dy_{1}^{\beta }\right) U_{,\alpha \beta }  \label{a4.19}
\end{equation}%
Using (\ref{a4.12}), (\ref{a4.14}), one obtains for (\ref{a4.17}) - (\ref%
{a4.19})%
\begin{eqnarray}
\delta _{1}U &=&-\frac{1}{2}c\mathbf{\beta \nabla }Udt+\frac{1}{8}c\mathbf{%
\beta \nabla }U\alpha dt+\frac{1}{4}c\mathbf{\dot{\beta}\nabla }U\left(
dt\right) ^{2}+\frac{c^{2}}{8}\beta ^{\alpha }\beta ^{\beta }U_{,\alpha
\beta }\left( dt\right) ^{2}  \label{a4.20} \\
\delta _{2}U &=&\frac{1}{2}c\mathbf{\beta \nabla }Udt+\frac{1}{8}c\mathbf{%
\beta \nabla }U\alpha dt+\frac{1}{4}c\mathbf{\dot{\beta}\nabla }U\left(
dt\right) ^{2}+\frac{c^{2}}{8}\beta ^{\alpha }\beta ^{\beta }U_{,\alpha
\beta }\left( dt\right) ^{2}  \label{a4.21}
\end{eqnarray}%
\begin{equation}
\delta _{2-1}U=\frac{1}{2}\left( \frac{1}{2}c\mathbf{\beta }\alpha dt+c%
\mathbf{\dot{\beta}}\left( dt\right) ^{2}\right) \mathbf{\nabla }U+O\left(
dt^{4}\right)  \label{a4.22}
\end{equation}

Using (\ref{a4.6}) and (\ref{a4.15}) (\ref{a4.16}), one obtains for the
infinitesimal quantity $\alpha $%
\begin{eqnarray}
\alpha &=&\frac{2\delta _{2}U-2\delta _{1}U+\mathbf{\beta }_{2}^{2}-\mathbf{%
\beta }_{1}^{2}}{1-2U-\mathbf{\beta }_{2}^{2}-2\delta _{2}U}=\frac{2\delta
_{2}U-2\delta _{1}U+2\mathbf{\beta \dot{\beta}}dt}{1-2U-\mathbf{\beta }^{2}}
\nonumber \\
&&+\frac{\left( 2\delta _{2}U-2\delta _{1}U+2\mathbf{\beta \dot{\beta}}%
dt\right) \left( \mathbf{\beta \dot{\beta}}dt+2\delta _{2}U\right) }{\left(
1-2U-\mathbf{\beta }^{2}\right) ^{2}}+O\left( dt^{3}\right)  \label{a4.23}
\end{eqnarray}

Let us substitute expansions (\ref{a4.12}) - (\ref{a4.23}) in the dynamic
equation (\ref{a4.5})/ We obtain after simplifications%
\begin{equation}
\frac{1}{2}\mathbf{\dot{\beta}}^{2}\left( dt\right) ^{2}-c\mathbf{\dot{\beta}%
\nabla }U\left( dt\right) ^{2}+\frac{1}{2}\frac{\left( c\mathbf{\beta \nabla 
}U+\mathbf{\beta \dot{\beta}}\right) ^{2}}{1-2U-\mathbf{\beta }^{2}}\left(
dt\right) ^{2}+\frac{c^{2}}{2}\beta ^{\alpha }\beta ^{\beta }U_{,\alpha
\beta }\left( dt\right) ^{2}=0  \label{a4.24}
\end{equation}%
Note, that the terms of the order of $dt$ disappear.

In terms of variables $\mathbf{v,\dot{v},}V$, defined by relations (\ref%
{a3.8a}), (\ref{a4.3}) the relation (\ref{a4.24}) has the form%
\begin{equation}
\frac{1}{2}\mathbf{\dot{v}}^{2}-\mathbf{\dot{v}\nabla }V+\frac{1}{2}\frac{%
\left( \mathbf{v\nabla }V+\mathbf{v\dot{v}}\right) ^{2}}{c^{2}\left(
1-2c^{-2}V-c^{-2}\mathbf{v}^{2}\right) }+\frac{1}{2c^{2}}v^{\alpha }v^{\beta
}V_{,\alpha \beta }=0  \label{a4.25}
\end{equation}

One obtains in the nonrelativistic approximation 
\begin{equation}
\frac{1}{2}\mathbf{\dot{v}}^{2}-\mathbf{\dot{v}\nabla }V=0  \label{a4.26}
\end{equation}%
It is evident, that one cannot determine three components of vector $\mathbf{%
\dot{v}}$ from one equation (\ref{a4.26}). One can determine only mean value 
$\left\langle \mathbf{\dot{v}}\right\rangle $ of vector $\mathbf{\dot{v}}$%
\textbf{, }choosing some principle of averaging.

Let us represent $\mathbf{v}$ in the form 
\begin{equation}
\mathbf{\dot{v}}=\mathbf{\dot{v}}_{\parallel }+\mathbf{\dot{v}}_{\perp
},\qquad \mathbf{\dot{v}}_{\parallel }=\mathbf{\nabla }V\frac{\left( \mathbf{%
\dot{v}\nabla }V\right) }{\left\vert \mathbf{\nabla }V\right\vert ^{2}}%
,\qquad \mathbf{\dot{v}}_{\perp }=\mathbf{\dot{v}-\nabla }V\frac{\left( 
\mathbf{\dot{v}\nabla }V\right) }{\left\vert \mathbf{\nabla }V\right\vert
^{2}}  \label{a4.27}
\end{equation}%
where $\mathbf{v}_{\parallel }$ and $\mathbf{v}_{\perp }$ are components of $%
\mathbf{v}$, which are parallel to $\mathbf{\nabla }V$ and perpendicular to $%
\mathbf{\nabla }V$ correspondingly. Let us suppose, that the mean value $%
\left\langle \mathbf{\dot{v}}\right\rangle $ of vector $\mathbf{\dot{v}}$ is
directed along the vector $\mathbf{\nabla }V.$ In this case $\left\langle 
\mathbf{\dot{v}}_{\perp }\right\rangle =0$, although $\left\langle \mathbf{%
\dot{v}}_{\perp }^{2}\right\rangle >0$, in general. One obtains from (\ref%
{a4.26}) 
\begin{equation}
\dot{v}_{\parallel }^{2}-2\dot{v}_{\parallel }\left\vert \mathbf{\nabla }%
V\right\vert +\left\langle \mathbf{\dot{v}}_{\perp }^{2}\right\rangle =0
\label{a4.28}
\end{equation}%
or 
\begin{equation}
\dot{v}_{\parallel }=\left\vert \mathbf{\nabla }V\right\vert \pm \sqrt{%
\left\vert \mathbf{\nabla }V\right\vert ^{2}-\left\langle \mathbf{\dot{v}}%
_{\perp }^{2}\right\rangle }  \label{a4.29}
\end{equation}%
It follows from (\ref{a4.29}), that%
\begin{equation}
0<\left\langle \mathbf{\dot{v}}_{\perp }^{2}\right\rangle \leq \left\vert 
\mathbf{\nabla }V\right\vert ^{2},\qquad 0<\dot{v}_{\parallel }<2\left\vert 
\mathbf{\nabla }V\right\vert  \label{a4.30}
\end{equation}%
At any admissible value $\left\langle \mathbf{\dot{v}}_{\perp
}^{2}\right\rangle $ the quantity $\dot{v}_{\parallel }$ vibrates around its
mean value $\left\langle \dot{v}_{\parallel }\right\rangle =\left\vert 
\mathbf{\nabla }V\right\vert $. Taking into account that $\left\langle 
\mathbf{\dot{v}}_{\perp }\right\rangle =0$, one obtains, that 
\begin{equation}
\left\langle \mathbf{\dot{v}}\right\rangle =\mathbf{\nabla }V=\mathbf{\nabla 
}\frac{GM}{r}  \label{a4.31}
\end{equation}

Any macroparticle moves in the gravitational field with the acceleration (%
\ref{a4.31}). Note, that for obtaining of this result only supposition $%
\left\langle \mathbf{\dot{v}}_{\perp }\right\rangle =0$ is important.
Variation of $\left\langle \mathbf{\dot{v}}_{\perp }^{2}\right\rangle $ do
not change the direction of the acceleration $\left\langle \mathbf{\dot{v}}%
\right\rangle .$According to (\ref{a4.29}) this variation can change only $%
\left\vert \dot{v}_{\parallel }\right\vert $, which can be compensated by a
proper change of the gravitational constant $G$.

In the nonrelativistic case the acceleration $\mathbf{\dot{v}}$ of a
macroparticle in the gravitational field is determined be the mean value $%
\left\langle \mathbf{\dot{v}}\right\rangle =\mathbf{\nabla }V$. This result
agrees with the Newtonian theory of gravitation.

In the general case we obtain instead of (\ref{a4.28}) 
\begin{eqnarray}
&&\dot{v}_{\parallel }^{2}-2\dot{v}_{\parallel }\left\vert \mathbf{\nabla }%
V\right\vert +\left\langle \mathbf{\dot{v}}_{\perp }^{2}\right\rangle 
\nonumber \\
&=&-\frac{\left( \mathbf{v\nabla }V\right) ^{2}+2\left( \mathbf{v\nabla }%
V\right) \left( \mathbf{v}\left\langle \mathbf{\dot{v}}_{\parallel
}\right\rangle \right) +\left\langle \left( v_{\mathbf{\parallel }}\dot{v}%
_{\parallel }+\mathbf{v}_{\perp }\mathbf{\dot{v}}_{\perp }\right)
^{2}\right\rangle }{c^{2}-2V-\mathbf{v}^{2}}-\frac{1}{c^{2}}v^{\alpha
}v^{\beta }V_{,\alpha \beta }  \label{a4.32}
\end{eqnarray}%
This result distinguishes from the conventional result of the general
relativity, because it depends on the second derivatives $V_{,\alpha \beta }$
of the gravitational potential. Equation (\ref{a4.32}) can be written in the
form of quadratic equation with respect to $\dot{v}_{\parallel }$ 
\begin{eqnarray}
&&\dot{v}_{\parallel }^{2}\left( 1+\frac{v_{\parallel }^{2}}{c^{2}-2V-%
\mathbf{v}^{2}}\right) -2\dot{v}_{\parallel }\left( \left\vert \mathbf{%
\nabla }V\right\vert -\frac{\left( \mathbf{v\nabla }V\right) v_{\parallel }}{%
c^{2}-2V-\mathbf{v}^{2}}\right) +\left\langle \mathbf{\dot{v}}_{\perp
}^{2}\right\rangle  \nonumber \\
&=&-\frac{\left( \mathbf{v\nabla }V\right) ^{2}+\left\langle \left( \mathbf{v%
}_{\perp }\mathbf{\dot{v}}_{\perp }\right) ^{2}\right\rangle }{c^{2}-2V-%
\mathbf{v}^{2}}-\frac{1}{c^{2}}v^{\alpha }v^{\beta }V_{,\alpha \beta }
\label{a4.33}
\end{eqnarray}

\section{Concluding remarks}

Our consideration of free particle motion led to the conclusion, that the
nonrelativistic motion of a free microparticle is multivariant (stochastic)
already in the gravitational field of a heavy sphere. It is multivariant for
other gravitational fields also, because the gravitational field has been
considered in the general form of gravitational potential. Free motion of a
macroparticle appears to be single-variant (deterministic), because
stochastic behavior of different microparticles inside the macroparticle is
independent. Averaging over many microparticles, one obtains a deterministic
free motion. In the nonrelativistic case this average motion of a
macroparticle coincides with predictions of the Newtonian theory of
gravitation, which have been tested experimentally. In the relativistic case
there are differences between predictions of the gravitational theory, based
on physical space-time geometry, and predictions of the contemporary
gravitational theory (general relativity), based on the Riemannian
space-time geometry.

The Riemannian space-time geometry is an approximate theory of the
space-time, because it is based on the \textit{false supposition, that the
Riemannian geometry is the most general geometry, which could be used for
the space-time description. }Besides, the Riemannian geometry is constructed
as a mathematical geometry, i.e. the geometry is considered as a logical
construction, but not as a science on mutual disposition of geometrical
objects. In particular, the space-time distance (world function) appears to
be many-valued even in the gravitational field of a heavy sphere. It is
inadmissible, if the space-time geometry is a physical geometry, i.e. a
science on mutual disposition of geometrical objects. The world function is
to be single-valued, as the main characteristic of the geometry.

The contemporary theory of gravitation admits one to determine only metric
tensor, which determines the world function \textit{only under supposition
on the Riemannian space-time geometry}. The new conception (generalization
of the general relativity on the case of a physical space-time geometry)
admits one to determine the world function directly, and this world function
appears to describe a non-Riemannian geometry even in the case of the
gravitational field of a heavy sphere. Generalization of the general
relativity appeared to be possible, only when we refuse a use of
nonrelativistic concepts in the relativity theory. In particular, the
nonrelativistic concept of nearness of two events has been replaced by the
relativistic one \cite{R2009}.

Interrelation between the multivariant physical space-time geometry and
conventional Riemannian geometry may be described as follows. The Riemannian
geometry is a single-variant approximation of the physical geometry,
generated by our poor knowledge of a geometry. From viewpoint of physical
geometry this approximation looks as follows. Multivariant world chains of
microparticles are replaced by averaged world chains, which are interpreted
as exact world lines of particles. The space-time geometry is constructed on
the basis of these "exact" world lines in such a way, that these world lines
are geodesics of the geometry. Such an approximation is possible at large
scales for description of the particle motion, although one cannot be sure,
that this approximation is effective at description of the matter
distribution influence on the space-time geometry, because the Riemannian
geometry is inconsistent. At small scales, when the multivariance of the
microparticle motion can be observed directly (for instance, diffraction of
electrons on small hole), one uses the quantum description, which takes into
account some properties of the multivariant motion.

The contemporary theory of gravitation, as well interpretation of
astronomical observations, based on this theory, needs a revision. In
particular, one needs to revise such concepts of the contemporary theory as
"black hole" and "dark matter". At the present stage of investigations, one
cannot state, that these concepts describe fictitious objects. However,
these concepts becomes disputable, as far as they are introduced on the
basis of a doubtful theory of gravitation. In any case a revision of the
contemporary gravitational theory is necessary.

\end{document}